\documentclass[graybox]{svmult}

\usepackage{type1cm}        % activate if the above 3 fonts are
\usepackage{makeidx}         % allows index generation
\usepackage{graphicx}        % standard LaTeX graphics tool
\usepackage{multicol}        % used for the two-column index
\usepackage[bottom]{footmisc}% places footnotes at page bottom

\usepackage{newtxtext}       % 
\usepackage[varvw]{newtxmath}       % selects Times Roman as basic font

\makeindex             % used for the subject index
\begin{document}

\title*{Gravitational-wave observations and primordial black holes}
\author{Francesco Iacovelli \and Michele Maggiore}
\institute{Francesco Iacovelli \email{Francesco.Iacovelli@unige.ch}
\\ Michele Maggiore \email{Michele.Maggiore@unige.ch} \at D\'epartement de Physique Th\'eorique, Universit\'e de Gen\`eve, 24 quai Ernest Ansermet, 1211 Gen\`eve 4, Switzerland, and Gravitational Wave Science Center (GWSC), Universit\'e de Gen\`eve, CH-1211 Geneva, Switzerland }

\maketitle

\abstract*{Gravitational-wave observations have the potential of allowing the identification of a population of merging primordial black-hole binaries. We provide an overview of the capabilities of present and future GW detectors, with a special emphasis on the perspective for observing quantities that are signatures of a primordial origin.}

\abstract{Gravitational-wave observations have the potential of allowing the identification of a population of merging primordial black-hole binaries. We provide an overview of the capabilities of present and future GW detectors, with a special emphasis on the perspective for observing quantities that are signatures of a primordial origin.}

\vspace{5mm}
In this chapter we discuss the perspectives for observing PBHs and studying their properties, both individually and as a population, with current and future GW detectors. 

In GW physics the last decade has been marked by the extraordinary results  of the LIGO-Virgo Collaboration. The first detection of  a BBH coalescence  took place  on Sept. 14th, 2015~\cite{LIGOScientific:2016aoc}, and was a milestone in science. Another   milestone was the observation, on Aug.~17, 2017, of the first binary neutron star coalescence~\cite{LIGOScientific:2017vwq}, and the detection and follow-up of the source in all bands of the electromagnetic spectrum~\cite{LIGOScientific:2017ync}. Nowadays,  after three observing runs, the catalog of observed compact binary coalescences  contains about 90 BBH coalescences,  two NSBH  binaries  and two BNS~\cite{LIGOScientific:2020ibl,KAGRA:2021vkt}. In the currently ongoing fourth observing run, one candidate event every few days  has been announced \cite{gracedb}. Many different fields have already been significantly impacted by these discoveries: we now have a much better  understanding of the population properties of compact binaries (see, e.g. ref.~\cite{KAGRA:2021duu}), we start to have first constraints on deviations from GR, as well as on the expansion history of the Universe (see, e.g. ref.~\cite{LIGOScientific:2021sio,LIGOScientific:2021aug}), while the 
multi-messenger observation of the binary neutron-star system GW170817 and of its electromagnetic counterpart \cite{LIGOScientific:2017ync,Monitor:2017mdv} had important implications in relativistic astrophysics, nuclear physics, nucleosynthesis in the Universe, and cosmology.

%In a different frequency band, corresponding to the nanoHz, another  exciting result came out in mid-2023 from pulsar timing arrays (PTAs),  that reported evidence for the presence of a stochastic background of GWs \cite{NANOGrav:2023gor,EPTA:2023fyk,Reardon:2023gzh,Xu:2023wog}, in decade-long data streams. While the origin of this signal has still to be understood, a  natural candidate is offered by the incoherent superposition of signal from the inspiral of supermassive BH binaries.  

While the  results of second-generation (2G) detectors such as  LIGO, Virgo and KAGRA have marked the beginning of the GW astronomy era, a new generation (3G) of ground-based detectors is currently under development. In particular, the European flagship project is 
Einstein Telescope (ET)~\cite{Hild:2008ng,Punturo:2010zz,Hild:2010id}, while 
the US community is developing  the  Cosmic Explorer  project~\cite{Reitze:2019iox,Evans:2021gyd}.
A  discussion of the ET Science Case can be found in \cite{Maggiore:2019uih}, while a recent comprehensive study  of the prospects for CBC observations and characterization at ET can be found in \cite{Iacovelli:2022bbs};  the perspective for multi-messenger observations with ET are discussed in detail in 
\cite{Ronchini:2022gwk}; for CE and for a more general discussion of 3G detectors  see also \cite{Borhanian:2022czq,Kalogera:2021bya}. More recently, the whole science case of ET has been refined and evaluated with respect to different possible choices of design, including different  options for the geometry (a single-site detector with nested interferometer in a triangular configuration, versus   two L-shaped interferometers) and different sensitivity curves~\cite{Branchesi:2023mws}.

\begin{figure}[t]
\sidecaption[t]
\includegraphics[scale=.45]{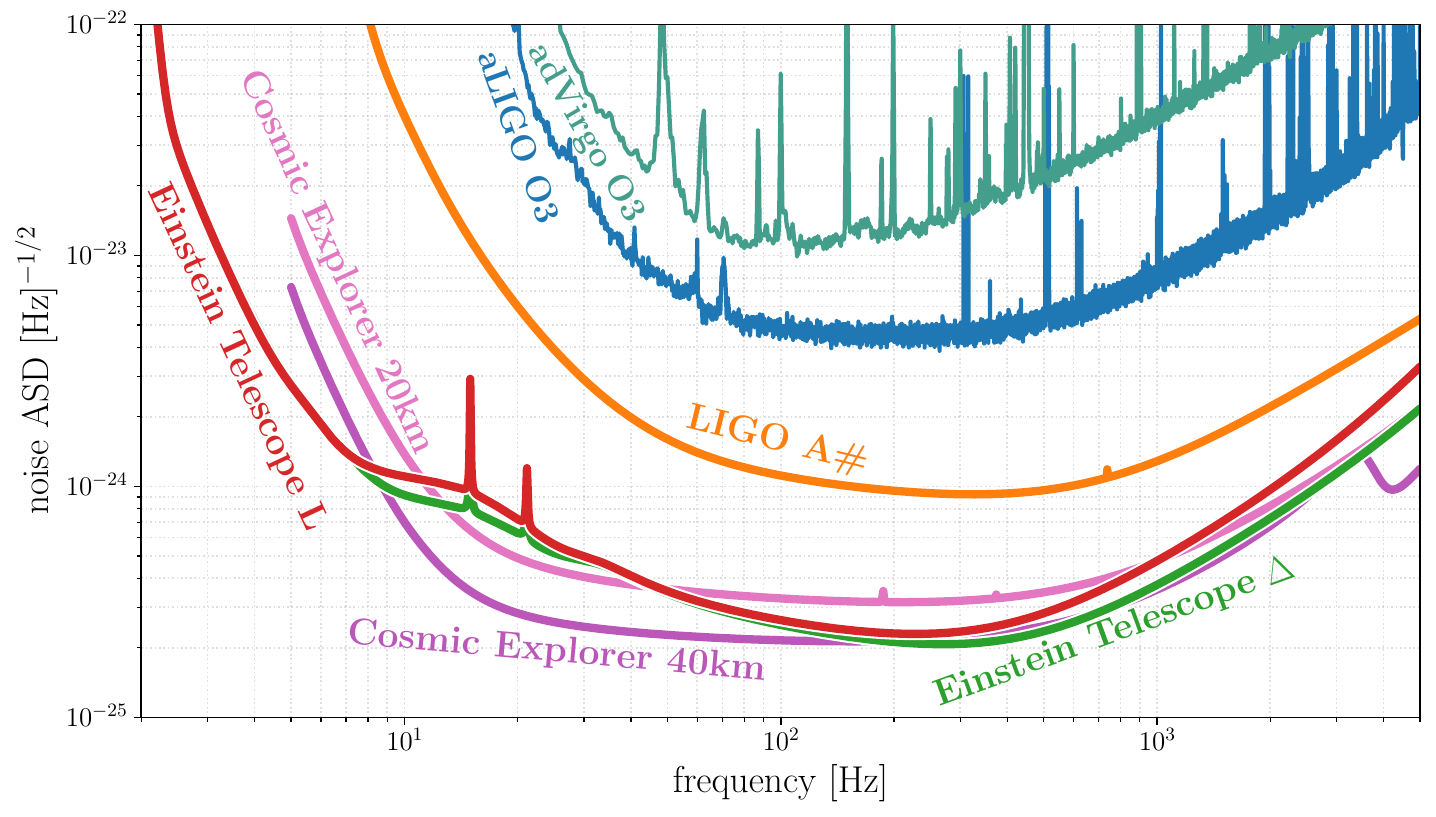}
\caption{Noise amplitude spectral densities for: the current second generation detectors LIGO and Virgo achieved during the O3 observing run; the prospects for LIGO in the so-called A\# configuration;  ET in the triangular configuration with 10~km  arms (rescaling the spectral density by a factor $2/3$ to take into account the three nested interferometers~\cite{Iacovelli:2022bbs});  a single  L-shaped interferometer (part of  a 2L configuration of ET) with 15~km arms; CE  with 20~km and with 40~km long arms.}
\label{fig:ASDs_GroundGWdets}
\end{figure}

\begin{figure}[t]
\sidecaption[t]
\includegraphics[scale=.45]{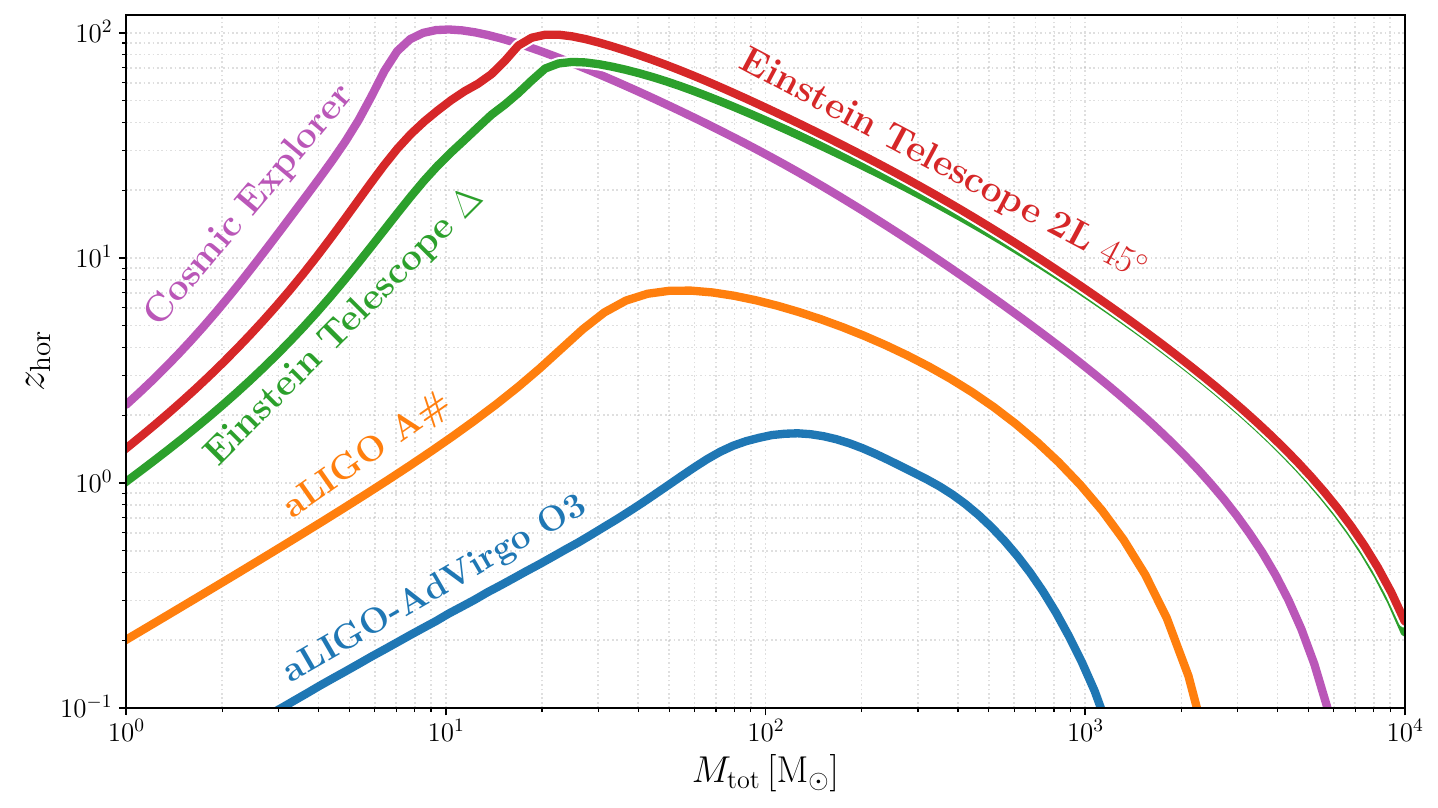}
\caption{Redshift horizons for equal-mass spinless binaries at: LIGO and Virgo during the O3 observing run; LIGO in the A\# configuration; ET both in the triangular configuration with 10~km  arms and as as two L-shaped detector with 15~km  arms at $45^{\circ}$; CE in a configuration consisting of two L-shaped instruments one with 40~km long arms and the other of 20~km long arms.}
\label{fig:Horizons_GroundGWdets}
\end{figure}

Fig.~\ref{fig:ASDs_GroundGWdets} shows the  strain sensitivities that were reached by LIGO and Virgo during the O3 run. As an example of the potential of the second-generation LIGO and Virgo instruments, when pushed to their limit, we show a representative sensitivity for the so-called A\# stage of the LIGO detectors. We compare these sensitivities with those expected for ET (in  the two  configurations currently under  consideration, i.e.  a triangle with 10~km arms and a 2L with 15~km arms) and CE (considering both a detector with 20~km arms and one with 40~km arms).
The corresponding detection horizons, for equal-mass spinless binaries, are shown in Fig.~\ref{fig:Horizons_GroundGWdets}.

From the point of view of PBHs the following aspects, in the jump form 2G to 3G detectors, are noteworthy.

1. The increased reach in distance. This is particularly significant for identifying a PBH population because, even if  the redshift at which the first stars formed is not precisely known, theoretical calculations and cosmological simulations suggest that  the redshift at which Pop~III remnant BBHs merge is below $z\sim 30$ (see \cite{Ng:2020qpk,Ng:2021sqn} and refs. therein). In contrast,  stellar-mass PBHs have a 
merger rate density that increases monotonically with redshift, up to $z\sim 10^3$ and beyond~\cite{Raidal:2018bbj}. While current detectors have a reach where astrophysical BH dominates, we see from the figure that 3G detectors reach redshift as high as $z\sim 50$ (and even $z\sim 100$ for the optimal combinations of masses and orientation), where the contamination from astrophysical BH binary mergers is expected to be negligible.

Reaching a great distance, however, is not yet sufficient to have a smoking-gun signature of a primordial origin. It is also necessary to be able to reconstruct the redshift with an error sufficiently small, so that the support of the posterior at $z$ smaller than a critical value, such as $z_c=30$, is negligible. This issue was studied in \cite{Ng:2021sqn,Ng:2022vbz,Mancarella:2023ehn}, where it is found that individual identification of PBHs based on the condition $z>30$ is possible but challenging, and in any case requires a network of 3G detectors including ET and one or two CE.

\begin{figure}[t]
\sidecaption[t]
\includegraphics[scale=.35]{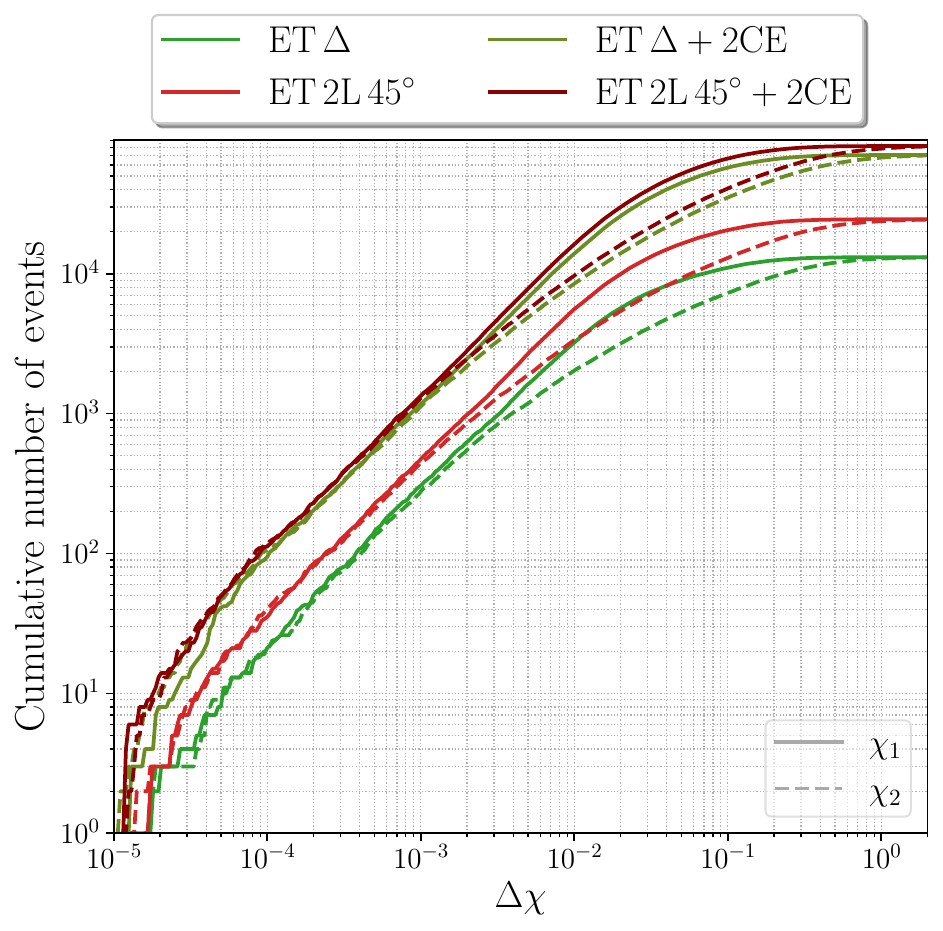}\ \includegraphics[scale=.35]{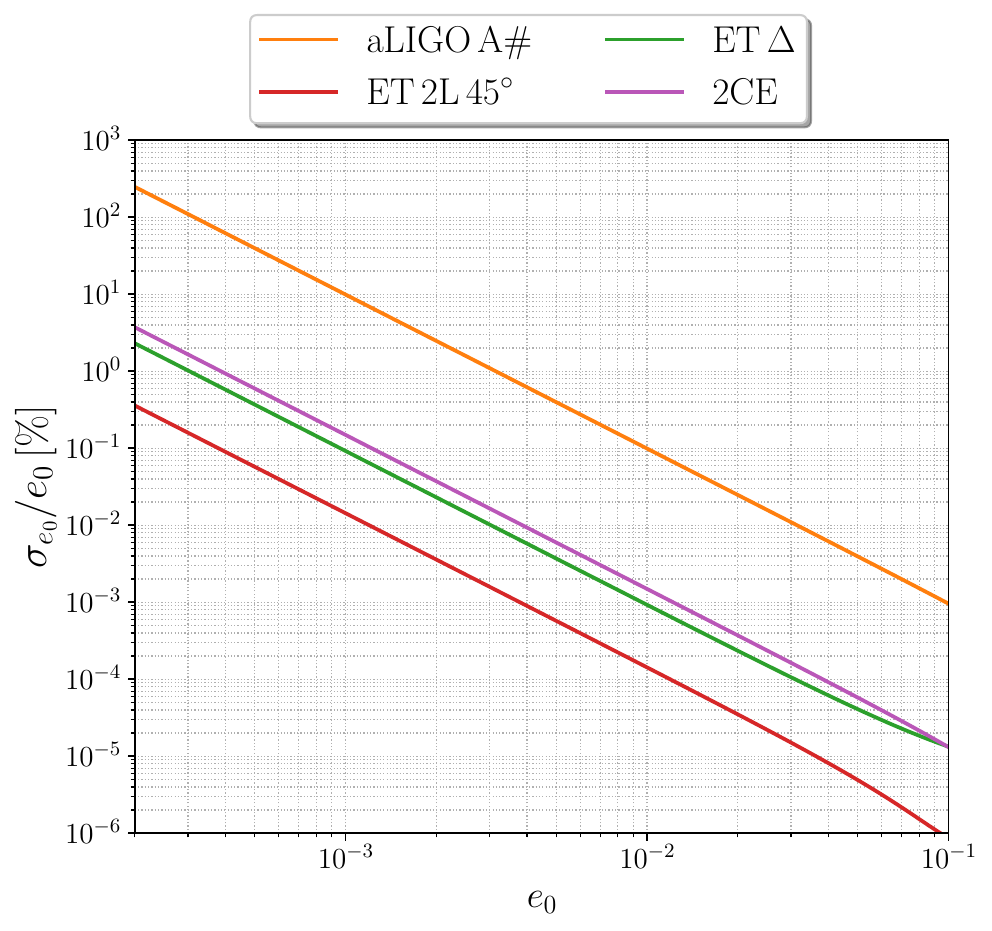}
\caption{\emph{Left panel:} cumulative distribution of the errors on the PBH binary spin parameters obtained with the \textsc{gwfast} Fisher code \cite{Iacovelli:2022bbs,Iacovelli:2022mbg} for the population used in \cite{Franciolini:2023opt} and based on \cite{Franciolini:2022tfm}. We report the results for two ET configurations (10~km triangle and 15~km 2L at $45^\circ$) both alone and in a network with 2 CE. \emph{Right panel:} Relative error on the eccentricity parameter $e_0$ at $f_{e_{0}}=10~{\rm Hz}$ for a $M_{\rm tot} = 20~{\rm M}_{\odot}$ spinless binary at 100~Mpc as a function of $e_0$. We compare the performance of aLIGO in the A$\#$ configuration, ET in the triangular and 2L configurations and 2 CE.}
\label{fig:spin_ecc_errs_GroundGWdets}
\end{figure}

\begin{figure}[t]
\sidecaption[t]
\includegraphics[scale=.45]{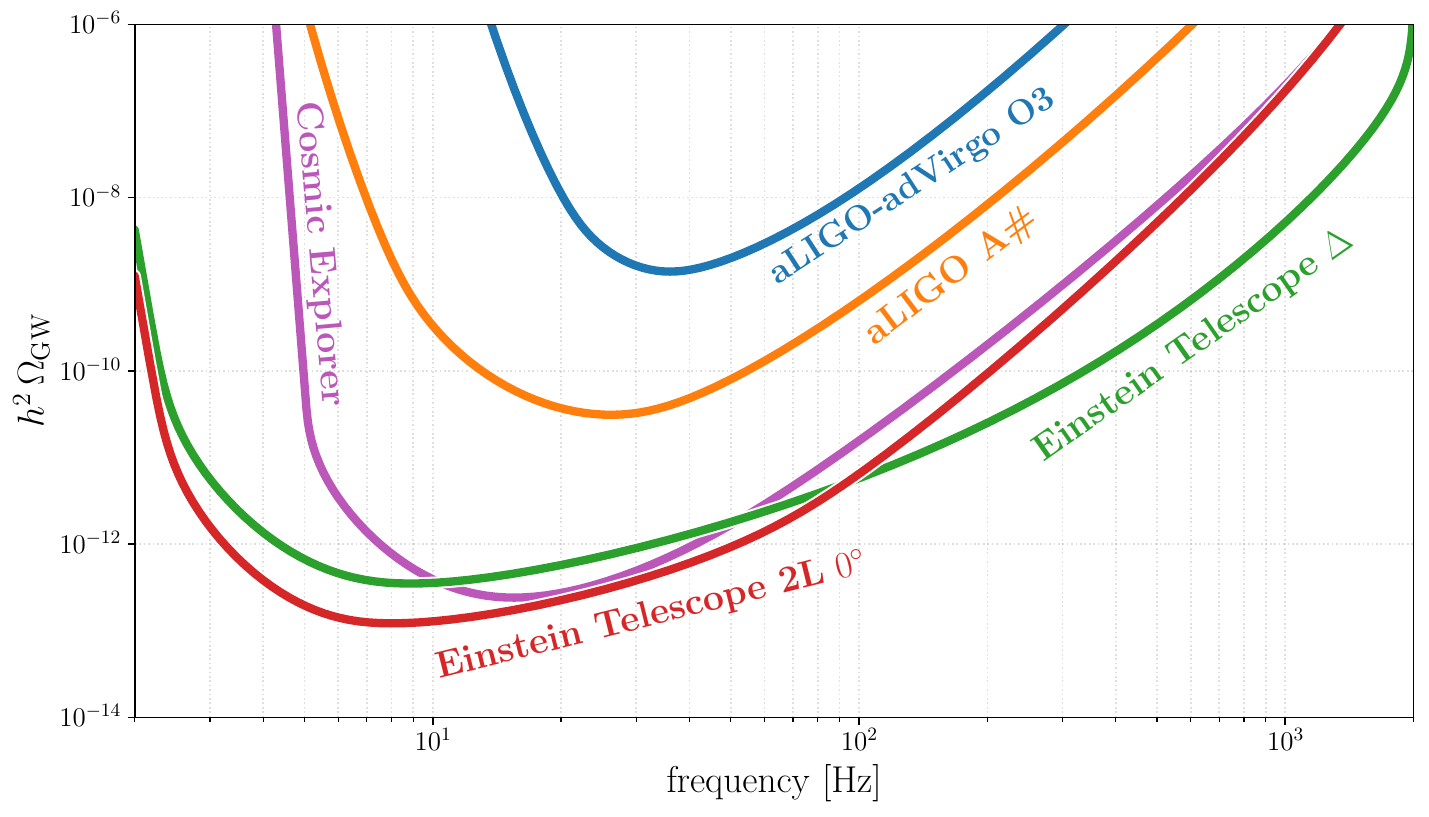}
\caption{Power-law integrated sensitivity curves to a stochastic background of GWs $\Omega_{\rm GW}$ \cite{Thrane:2013oya} obtained for: LIGO and Virgo during the O3 observing run; LIGO in the A\# configuration; ET both in the triangular configuration with 10~km  arms and as as two parallel L-shaped detector with 15~km arms; CE in a configuration consisting of two L-shaped instruments one with 40~km long arms and the other of 20~km long arms. The curves are scaled by $h=H_0/100~{\rm km\,s}^{-1}\,{\rm Mpc}^{-1}$ and obtained assuming an observation time of 1~yr and an SNR of 1.}
\label{fig:PLSs_GroundGWdets}
\end{figure}

\vspace{2mm}
2. The significant improvement in sensitivity of 3G detectors implies a corresponding improvement in the parameter estimation of coalescing binaries. From the point of view of PBHs, particularly relevant are the orbit eccentricity, and the  spins of the binary components \cite{Franciolini:2021xbq}. The left panel of Fig.~\ref{fig:spin_ecc_errs_GroundGWdets} shows the improvement in the reconstruction of $\chi_{1,z}$ and  $\chi_{2,z}$, the components of the spin parallel to the orbital angular momentum, for the two component stars (we use the label 1 to refer to the heavier star and 2 to the lighter star), assuming a population of PBHs as in \cite{Franciolini:2023opt} based on the results in \cite{Franciolini:2022tfm}, while the right panel shows the accuracy on the eccentricity for a source at a fixed distance $d_L=100$~Mpc and total source-frame mass $20 {\rm M}_{\odot}$.

\vspace{2mm}
3. A possible smoking gun signature for PBH is a source-frame mass in the subsolar range. Fig.~\ref{fig:Horizons_GroundGWdets} shows that, in this regime, even LIGO-A\# cannot go significantly beyond our local Universe. In contrast, ET or CE can reach cosmological distances, up to $z\sim 1-2$ for a total mass of $1  {\rm M}_{\odot}$ and $z\sim 2-5$ for a total mass of $2  {\rm M}_{\odot}$.

Finally, Fig.~\ref{fig:PLSs_GroundGWdets} show the sensitivity to stochastic background energy fraction $\Omega_{\rm GW}$, as expressed by the  power-law integrated sensitivity curve~\cite{Thrane:2013oya}. We see that 3G detector gain two orders of magnitudes even with respect to LIGO-A\#.
The consequences for PBHs will be developed in a different section

%Similar figures of merit to the ones reported above for the LISA detector can be found e.g. in \cite{LISA:2017pwj, LISACosmologyWorkingGroup:2022jok}
\vspace{2mm}
The future will also be marked by pioneering experiments in completely different frequency bands as compared to the ones mentioned above, providing a new perspective on the universe. 

\begin{figure}[t]
\sidecaption[t]
\includegraphics[scale=.32]{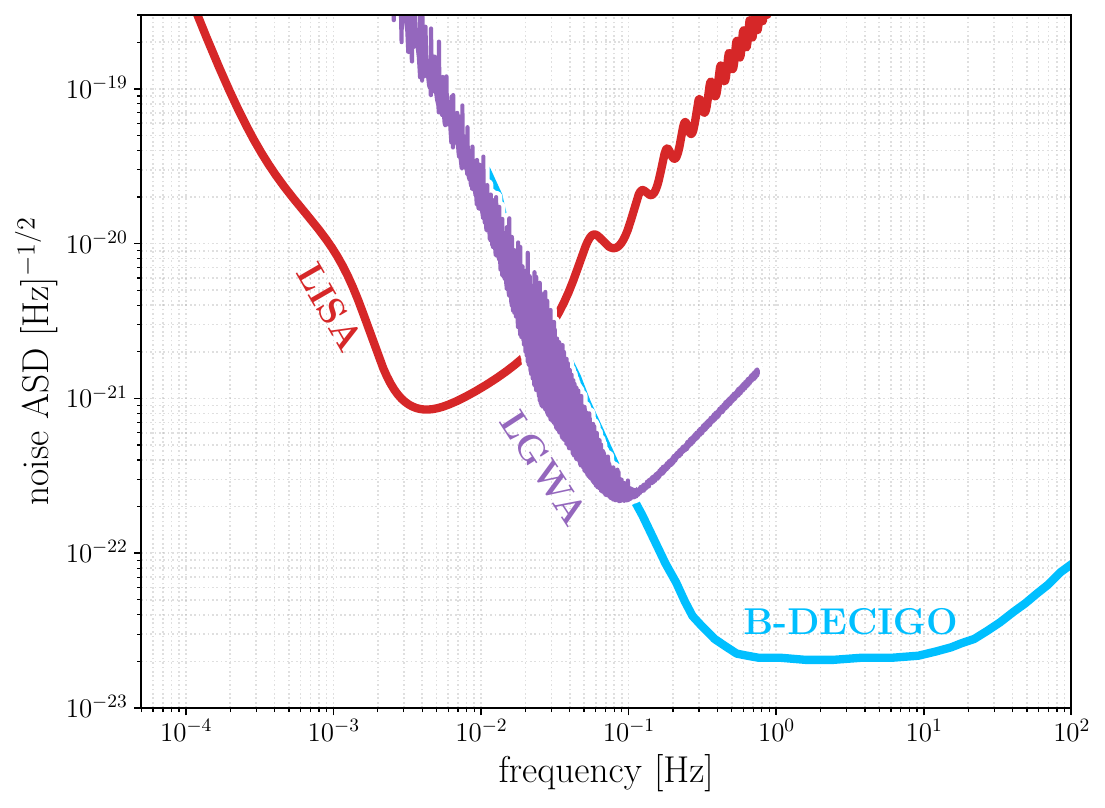}\ \includegraphics[scale=.32]{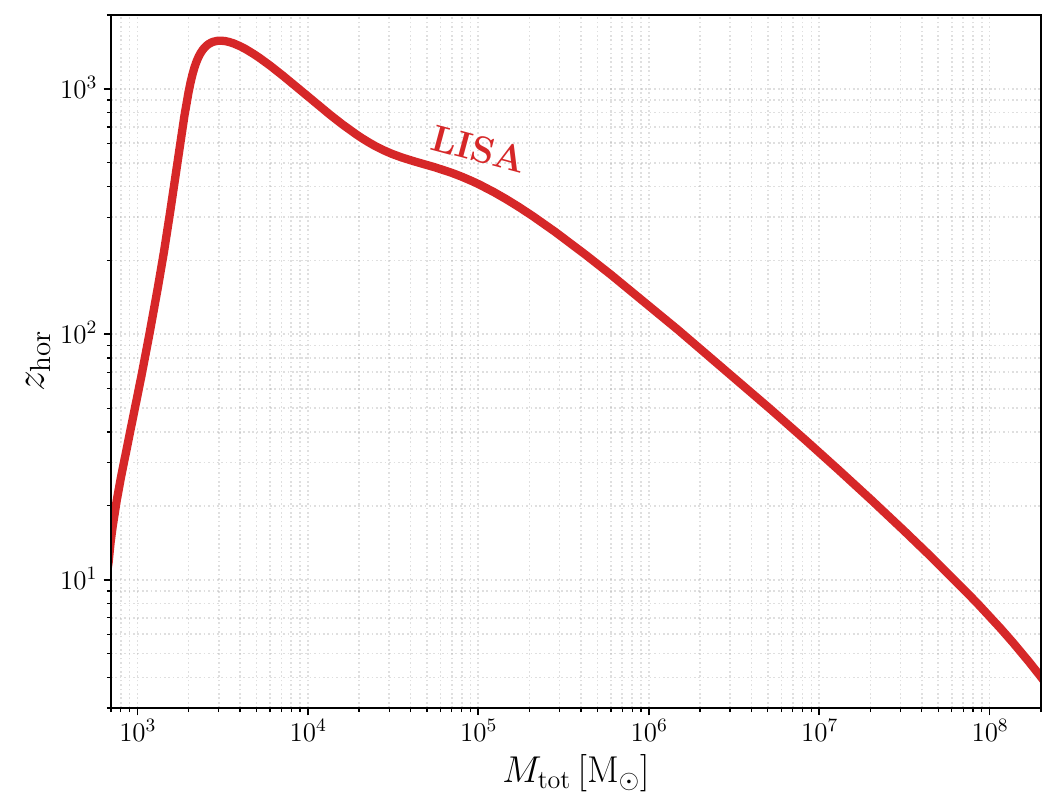}
\caption{\emph{Left panel:} Noise amplitude spectral densities for: LISA assuming an arm length of $2.5\times10^9\,{\rm km}$ and two data channels \cite{Robson:2018ifk}; LGWA \cite{LGWA:2020mma,Dupletsa:2022scg}; B-DECIGO \cite{Nakamura:2016hna,Kawamura:2020pcg}. \emph{Right panel:} Redshift horizon for equal-mass spinless binaries at LISA assuming a mission lifetime of 4 years.}
\label{fig:ASDs_Hor_LISA}
\end{figure}

\vspace{2mm}
The milliHz band will be explored by the space-based European Space Agency experiment LISA \cite{LISA:2017pwj} that with its three spacecraft arranged in a triangular configuration at million-km distance could collect signals from a large variety of sources: galactic binaries observations will allow to map the Milky Way through gravitational waves, giving valuable information on the formation of such systems and their evolution through gravitational and tidal dissipation; observing the inspiral and merger of massive and supermassive BHs through  cosmic time will help shedding light on their origin and evolution, and also allow us to put constraints on the expansion history of the Universe; detections of extreme mass ratio inspirals of stellar-origin BHs into massive BHs at the center of galaxies are an extremely promising way to map with exquisite precision the gravitational field around these gargantuan objects and study their environments; multiband of stellar-origin BHs in LISA and ground-based detectors would complement each other, giving e.g. the possibility to disentangle formation channels thanks to eccentricity measurements; the detection of stochastic backgrounds of cosmological origin generated that can be generated by several processes in the early universe can give extremely valuable information on the first instants after the Big Bang \cite{LISACosmologyWorkingGroup:2022jok}. %from the inspiral and merger of supermassive BHs, as well as a large variety of different sources, e.g. galactic WD binaries, to map the Milky Way through GWs, EMRIs, who could be used to map with exquisite precision the gravitational field around massive BHs, and stochastic backgrounds of cosmological origin generated by several processes in the early universe \cite{LISACosmologyWorkingGroup:2022jok}. 
In Fig.~\ref{fig:ASDs_Hor_LISA} we report the sensitivity curve (left panel) and detection horizon for equal-mass spinless binaries (right panel) for the LISA detector, again noticing the considerable reach in terms of distance in a different mass range, giving e.g. the possibility to observe PBH binaries in a broad mass range out to extremely high redshift. 

\vspace{2mm}
The frequency gap between LISA and ground-based detectors such as ET and CE could be covered by proposed experiments such as the Lunar Gravitational Wave Antenna (LGWA) \cite{LGWA:2020mma}, that aims at using the Moon as a GW-detector, monitoring with seismometers the vibrations of the Moon; by experiments based on atom interferometry such as MIGA~\cite{Canuel2018}, VLBAI~\cite{schlippert2020matter} and  AION~\cite{Badurina:2019hst} (see also \cite{Proceedings:2023mkp} for review); and by the space interferometer DECIGO currently under study \cite{Kawamura:2020pcg}. As an example, the sensitivity of LGWA and the so-called B-DECIGO are reported in Fig.~\ref{fig:ASDs_Hor_LISA}.

\begin{figure}[t]
\sidecaption[t]
\includegraphics[scale=.45]{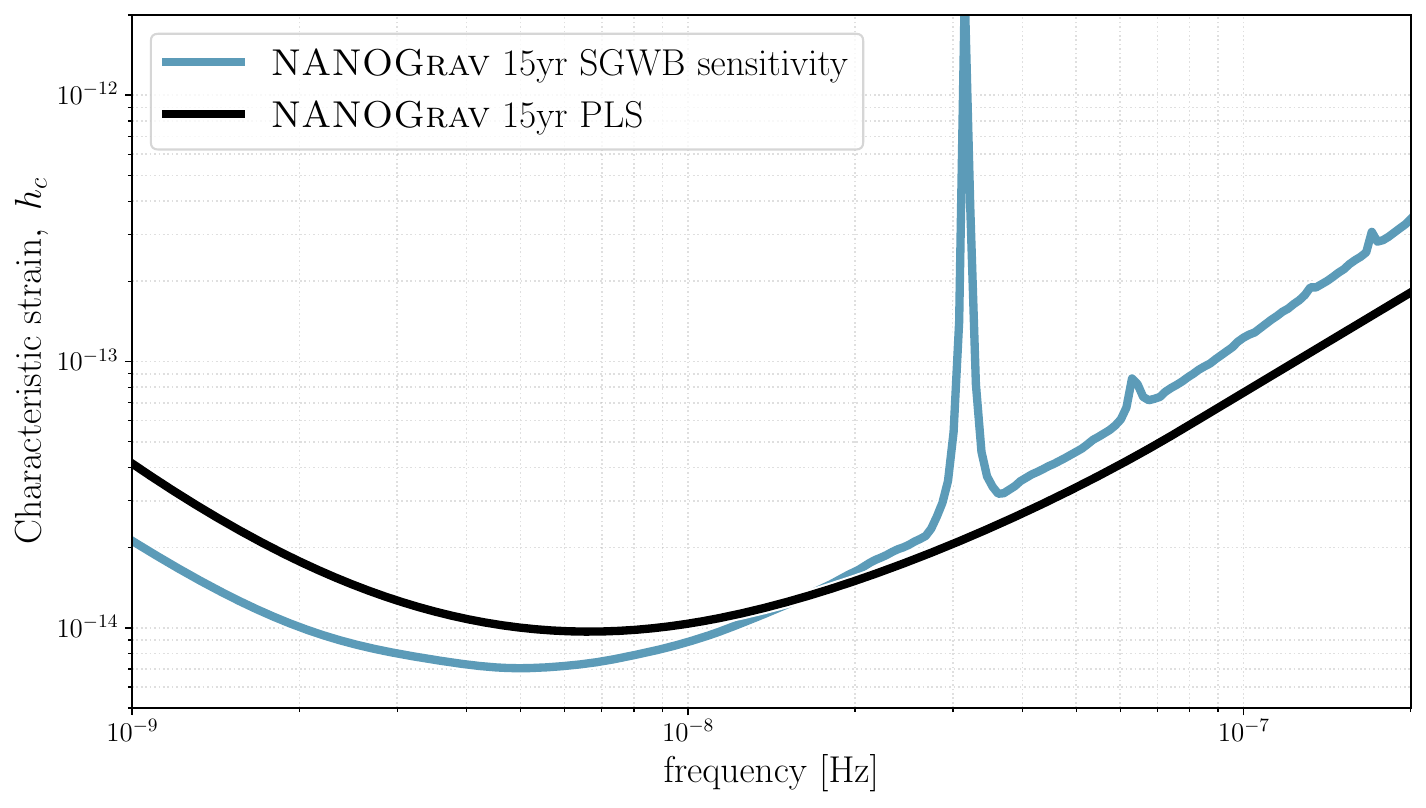}
\caption{Effective characteristic strain noise to a stochastic GW background source for the \textsc{NANOGrav} 15-yr PTA data and corresponding power-law integrated sensitivity curve \cite{Thrane:2013oya} obtained as in \cite{NANOGrav:2023ctt,Hazboun:2019vhv}.}
\label{fig:NANOGrav_sensitivity}
\end{figure}

\vspace{2mm}
In a different frequency band, corresponding to the nanoHz, another  exciting result came out in mid-2023 from pulsar timing arrays (PTAs): several experiments monitoring the arrival times of pulses emitted by galactic millisecond pulsars and searching for the correlation induced by GWs, such as the European Pulsar Timing Array (EPTA) and \textsc{NANOGrav} in North America, reported evidence for the presence of a stochastic background of GWs \cite{NANOGrav:2023gor,EPTA:2023fyk,Reardon:2023gzh,Xu:2023wog}, in data streams spanning longer than a decade. While the origin of this signal has still to be understood, a  natural candidate is offered by the incoherent superposition of signals from the inspiral of supermassive BH binaries. In Fig.~\ref{fig:NANOGrav_sensitivity} we report the effective characteristic strain noise for the data-release of 15 years of observations from the \textsc{NANOGrav} collaboration in the search for a stochastic signal, and the corresponding power-law integrated sensitivity curve. Thanks to more data being collected and new facilities with outstanding accuracy expected to become available, such as the Square Kilometre Array (SKA), the accuracy of PTA experiments is expected to greatly raise in the future. 

\bibliographystyle{unsrt}
\bibliography{mybib.bib}

\end{document}